\documentclass{tran-l}
\usepackage{amsmath, amssymb}
\usepackage{graphicx}
\usepackage{color}
\usepackage{textcomp}

\usepackage{graphics}

 \newtheorem{thm}{Theorem}[subsection]

 \theoremstyle{definition}
 
 \theoremstyle{remark}
 \newtheorem{rem}[thm]{Remark}
 \numberwithin{equation}{subsection}
 \newtheorem{exa}[thm]{Example}
 \numberwithin{equation}{section}
\newcommand{\bm}{\bibitem}

\newcommand{\be}{\begin{equation}}
\newcommand{\ee}{\end{equation}}

\newcommand{\bea}{\begin{eqnarray}}
\newcommand{\eea}{\end{eqnarray}}
\newcommand{\beaa}{\begin{eqnarray*}}
\newcommand{\eeaa}{\end{eqnarray*}}

\newcommand{\NN}{{\mathbb N}}

\newcommand{\RR}{{\mathbb R}}
\begin{document}

\title[Supersymmetric Methods in
Option Pricing]{Solvable Local and Stochastic Volatility Models:
Supersymmetric Methods in Option Pricing}

\author{Pierre Henry-Labord\`ere}
\email{phl@hotmail.com}

%%% ----------------------------------------------------------------------
\keywords{Solvable diffusion process, supersymmetry, differential
geometry}

%%% ----------------------------------------------------------------------
\maketitle
%%% ----------------------------------------------------------------------

\begin{abstract}
\noindent In this paper we provide an extensive classification of
one and two dimensional diffusion processes which admit an exact
solution to the Kolmogorov (and hence Black-Scholes) equation (in
terms of hypergeometric functions). By identifying the
one-dimensional solvable processes with the class of {\it
integrable superpotentials} introduced recently in supersymmetric
quantum mechanics, we obtain new analytical solutions. In
particular, by applying {\it supersymmetric transformations} on a
known solvable diffusion process (such as the Natanzon process for
which the solution is given by a hypergeometric function), we
obtain a hierarchy of new solutions. These solutions are given by
a sum of hypergeometric functions, generalizing the results
obtained in the paper "Black-Scholes Goes Hypergeometric"
\cite{alb}. For two-dimensional processes, more precisely
stochastic volatility models, the classification is achieved for a
specific class called gauge-free models including the Heston
model, the $3/2$-model and the geometric Brownian model. We then
present a new exact stochastic volatility model belonging to this
class.
\end{abstract}

%%% ----------------------------------------------------------------------

\section{Introduction}

For most mathematical models of asset dynamics, an exact solution
for the corresponding Kolmogorov \& Black-Scholes equation is
usually not available: there are, however, a few notable
exceptions. The known solutions for local volatility models are
the constant elastic of variance (CEV) \cite{cox1} including the
classical log-normal Black-Scholes process. For the instantaneous
short rate models, there are the CIR process \cite{cox2} (Bessel
process) and the Vacicek-Hull-White process \cite{hul}
(Ornstein-Uhlenbeck process). For stochastic volatility models,
the known exact solutions are the Heston model \cite{hes}, the
$3/2$-model and the geometric Brownian model \cite{lew}. These
analytical solutions can be used for calibrating a model quickly
and efficiently or can serve as a benchmark for testing the
implementation of more realistic models requiring intensive
numerical computation (Monte-Carlo, partial differential
equation). For example, the existence of a closed-form solution
for the price of a vanilla option in the Heston model allows us to
quickly calibrate  the model to the implied volatilities observed
on the market. The calibrated model can then be used to value
path-dependent exotic options using, for example, a Monte-Carlo
methodology.

\noindent In this paper, we show how to obtain new analytic
solutions to the Kolmogorov \& Black-Scholes equation, which we
refer to as  KBS throughout the rest of the paper, for 1d \& 2d
diffusion processes. In order to get to our classification, we
first present  a general reduction method to simplify the
multi-dimensional KBS equation. Rewriting the KBS equation as a
heat kernel equation on a Riemannian manifold endowed with an
Abelian connection, we show that this covariant equation can be
simplified using both the group of diffeomorphisms (i.e. change of
variables) and the group of Abelian gauge transformations. In
particular for the models admitting a {\it flat Abelian
connection}, there always exists a gauge transformation that
eliminates the Abelian connection of the diffusion operator.

\noindent In the second part we apply the reduction method,
previously presented, to a one-dimensional, time-homogeneous
diffusion process. Modulo a change of variable, the metric becomes
flat and the Abelian connection is an exact one-form for which a
gauge transformation can always be applied. Using these two
transformations, the resulting KBS equation becomes a {\it
Euclidean Schrodinger equation} with a scalar potential. Extensive
work has already been done to classify the set of scalar
potentials which admit an exact solution. In particular, using a
supersymmetric formulation of the Schrodinger equation which
consists in doubling the BKS equation with another equation, we
show how to generate a hierarchy of new solvable diffusion
processes starting from a solvable diffusion process (for e.g. a
Natanzon potential \cite{nat}). In this context, the local
volatility function is identified with a superpotential. Applying
supersymmetric transformations on the Natanzon potential (which is
the most general potential for which the Schrodinger equation can
be reduced to either a hypergeometric or a confluent equation), we
obtain a new class of solvable one-dimensional diffusion processes
which are characterized by six parameters.

\noindent The classification of one-dimensional time-homogeneous
solvable diffusion processes for which the solution to the KBS
equation can be written as a hypergeometric function has been
achieved in \cite{alb,alb1,alb2,alb3,kuz} using the well known
Natanzon classification. The application of supersymmetric
techniques to the classification of solvable potentials for the
Schrodinger equation has been reviewed in \cite{coo} where a large
number of references can be found. For the Kolmogorov \&
Fokker-Planck equations, one can consult \cite{kry,jun}.

\noindent In the last part we pursue this classification for
stochastic volatility models which admit a flat Abelian
connection: we refer to these as gauge-free models. Surprisingly,
this class includes all the well known exact stochastic volatility
models (i.e. the Heston model, the $3/2$-model and the geometric
Brownian model). For this gauge-free models, we reduce the
two-dimensional KBS equation to a Euclidean Schrodinger equation
with a scalar potential. Then, we present a new exact stochastic
volatility model which is a combination of the Heston and
$3/2$-models.

\section{Reduction method}
\noindent In this section, we explain how to simplify the KBS
equation. This reduction method will be used in the next section
to classify the solvable one and two dimensional time-homogeneous
processes. This method  is already well known for a
one-dimensional process and is presented in \cite{car,lip,lin} for
example. However, the extension of this method to
multi-dimensional diffusion processes requires the introduction of
differential geometric objects such as a metric and an Abelian
connection on a Riemannian manifold, as we will presently explain.

\noindent Let us assume that our time-homogeneous
multi-dimensional diffusion model depends on $n$ random processes
\noindent $x_i$ which can be either traded assets or hidden Markov
processes (such as a stochastic volatility $a$ or an instantaneous
short rate $r$). Let us denote $x=(x_i)_{i=1,\cdots,n}$, with
initial conditions $\alpha=(\alpha_i)_{i=1,\cdots,n}$. These
variables $x_i$ satisfy the following stochastic differential
equations (SDE)

\bea &&dx^i =b^i(x)dt+\sigma^i(x) dW_i \label{se}\\
 &&dW_idW_j=\rho_{ij}(x)dt \eea

\noindent with the initial condition $x_i(t=0)=\alpha_i$. The
no-arbitrage condition implies that there exists an equivalent
measure ${\mathbb P}$ such that the traded assets are (local)
martingales under this measure. For ${\mathbb P}$, the drifts
$b_i$ are consequently zero for the traded assets. Note that the
measure ${\mathbb P}$ is not unique as the market is not
necessarily complete. Finally, the fair value of a (European)
option, with payoff $f(x^i)$ at maturity $T$, is given by the
discounted mean value of the payoff $f$ conditional on the
filtration ${\mathcal F}_t$ generated by the Brownian motions $\{
W^i_{s \leq t} \}$

\bea {\mathcal C}(\alpha,t,T)={\mathbb E}^{\mathbb P}[
e^{-\int_t^T r_s ds }f |{\mathcal F}_t] \label{fair2} \eea

\noindent with $r_s$ the instantaneous short rate. This mean-value
depends on the probability density $p(x_{i},T|\alpha,t)$ which
satisfies the backward Kolmogorov equation ($\tau=T-t$,
$\partial_i={\partial \over \partial x_i}$)

\bea {\partial p \over \partial \tau}=b^i{\partial_i p}+{1 \over
2}\rho_{ij}\sigma^i\sigma^j{\partial_{ij} p } \label{fp1} \eea

\noindent with initial condition $p(\tau=0)=\delta(x-\alpha)$. In
this equation we have used the Einstein convention meaning that
two repeated indices are summed.

\noindent Using the Feymann-Kac theorem, one can show that the
fair value ${\mathcal C}$ for the option satisfies the
Black-Scholes equation ($\partial_i={\partial \over \partial
\alpha_i}$)

\bea {\partial {\mathcal C} \over \partial \tau}=b^i{\partial_i
{\mathcal C}}+g^{ij}{\partial_{ij}{\mathcal C}} -r(\alpha)
{\mathcal C} \label{fp2} \eea

\noindent with initial condition ${\mathcal
C}(\tau=0,\alpha)=f(\alpha)$.

\noindent In the following, the partial differential equations
(PDE) (\ref{fp1},\ref{fp2}) will be interpreted as the heat kernel
on a general smooth {$n$-dimensional manifold} $M$ (without a
boundary) endowed with a {metric} $g_{ij}$. The inverse of the
metric $g^{ij}$ is defined by \bea g^{ij}={1 \over 2}\rho_{ij}
\sigma^i \sigma^j \label{inversemetric}\eea \noindent and the
metric ($\rho^{ij}$ inverse of $\rho_{ij}$)

\bea g_{ij}=2 {\rho^{ij} \over \sigma^i \sigma^j} \label{metric}
\eea \noindent The differential operator \bea D=b^i\partial_i +
g^{ij}\partial_{ij} \label{d2}\eea which appears in (\ref{fp1}) is
a second order elliptic operator of Laplace type. We can then show
that there is a unique {connection} $\nabla$ on ${\mathcal L}$, a
{line bundle} over $M$, and a unique smooth function $Q$ on $M$
such that \bea D&\equiv&g^{ij} \nabla_i \nabla_j+Q
\\&=&g^{-{1 \over 2}}(\partial_i +{\mathcal A}_i) g^{1 \over
2}g^{ij}(\partial_j +{\mathcal A}_j) +Q \label{d1}\eea

\noindent with $g=det[g_{ij}]$. We may express the connection
${\mathcal A}^i$ and $Q$ as a function of the drift $b_i$ and the
metric $g_{ij}$ by identifying in (\ref{d1}) the terms
$\partial_i$ and $\partial_{ij}$ with those in (\ref{d2}). We find

\bea {\mathcal A}^i&=&{1 \over 2}(b^i-g^{-{1 \over
2}}\partial_j(g^{1/2}g^{ij})) \label{aaa} \\
Q&=&g^{ij}({\mathcal A}_i{\mathcal A}_j-b_j {\mathcal
A}_i-\partial_j {\mathcal A}_i) \label{qqq}\eea

\noindent Note that the Latin indices $i$,$j \cdots$ can be
lowered or raised using the metric $g_{ij}$ or its inverse
$g^{ij}$. For example ${\mathcal A}_i=g_{ij} {\mathcal A}^j={1
\over 2}(g_{ik}b^k-{1 \over 2}\partial_i ln(g) -g_{ip}\partial_k
g^{pk})$.

\noindent Using this connection, (\ref{fp1}) can be rewritten in
the covariant way, i.e.

\bea {\partial  \over \partial \tau}p(x,\alpha,\tau)=D
p(x,\alpha,\tau) \label{hk1}\eea

\noindent If we take ${\mathcal A}_i=0 \;, Q=0$ then $D$ becomes
the Laplace-Beltrami operator (or Laplacian) $\Delta=g^{-{1 \over
2}}\partial_ig^{1 \over 2}g^{ij}\partial_j$. For this
configuration, (\ref{hk1}) will be called the Laplacian heat
kernel. Similarly, the Black-Scholes equation (\ref{fp2}) can be
rewritten

\bea {\partial  \over \partial \tau}{\mathcal
C}(\alpha,\tau)=(D-r) {\mathcal C}(\alpha,\tau) \label{hk2}\eea

\vskip 3truemm

\noindent The Heat kernel equation can now be simplified by
applying the actions of the following groups.

\vskip 3truemm

\noindent $\rhd$ The group of diffeomorphisms $Diff({\mathcal M})$
which acts on the metric $g_{ij}$ and the connection ${\mathcal
A}_i$ by

\bea
f^* g_{ij}&=&g_{pk} \partial_i f^p(x) \partial_j f^k(x) \\
f^* {\mathcal A}_{i}&=&{\mathcal A}_{p} \partial_i f^p(x)  \; , \;
f \in Diff({\mathcal M})\eea

\noindent $\rhd$ The group of gauge transformations $\mathcal G$
which acts on the conditional probability (and the fair value
${\mathcal C}$) by \bea
p'&=&e^{\chi(x,\tau)-\chi(x=\alpha,\tau=0)}p
\\
{\mathcal C}'&=&e^{\chi(\alpha,\tau)}{\mathcal C} \label{gt} \eea

\noindent Then $p'$ (${\mathcal C}'$) satisfies the same equation
as $p$ (${\mathcal C}$) (\ref{hk1}) only with \bea {\mathcal A}'_i
&\equiv& {\mathcal
A}_i-\partial_i \chi \label{gc}\\
{Q}' &\equiv& Q+\partial_\tau \chi \eea

\noindent The constant phase $e^{\chi(x=\alpha,\tau=0)}$ has been
added so that $p$ and $p'$ satisfy the same boundary condition at
$\tau=0$. The above transformation is called a {\it gauge
transformation}. If the connection $\mathcal A$ is an exact form
(meaning that there exists a smooth function $\Lambda$ such that
${\mathcal A}_i=\partial_i \Lambda$), then by applying a gauge
transformation, we can eliminate the connection so that the heat
kernel equation for $p'$ (or ${\mathcal C}'$) has a  connection
equal to zero. \noindent It can be shown that for a
simply-connected manifold, the statement "${\mathcal A}$ is exact"
is equivalent to ${\mathcal F}=0$, where $\mathcal F$ is the
2-form curvature given in  a specific coordinate system by \bea
{\mathcal F}_{ij}=\partial_i {\mathcal A}_j -
\partial_j {\mathcal A}_i \eea
\noindent In the following, we will restrict our classification to
those processes for which ${\mathcal F}=0$, meaning there exists a
gauge transformation such that the transformed connection
vanishes. The operator $D$ reduces in this case to the symmetric
operator $D=\Delta+Q$ for which we can use an eigenvector
expansion.

\vskip 3truemm

\noindent {\bf Spectral decomposition}

\vskip 3truemm

\noindent This spectral decomposition is valid only if the
symmetric operator $D$ is an (unbounded) self-adjoint operator or
admits a self-adjoint extension. This will depend strongly on the
boundary conditions. In order to show that $D$ is self-adjoint or
admits self-adjoint extensions, we can use the {\it deficiency
indices} technique introduced by Von Neumann  \footnote{Before
explaining this technique, let us recall some definitions.

\noindent An operator $(H,{\mathcal D}(H))$ defined on a Hilbert
space ${\mathcal H}$ is said to be {\it densely defined} if the
subset ${\mathcal D}(H)$ is dense in ${\mathcal H}$, i.e.  for any
$\phi \in {\mathcal H}$ one can find in ${\mathcal D}(H)$ a
sequence which converges in norm to $\phi$.

\noindent The {\it domain} ${\mathcal D}(H^\dag)$ of an adjoint
operator of an (unbounded) operator $H$ with dense domain
${\mathcal D}$ is the space of functions $\psi$ such that the
linear form $\phi \rightarrow (\psi,H \phi)$ is continuous for the
norm of ${\mathcal H}$. Hence using Riesz' theorem, there exists a
unique $\phi'$ such that $(\psi,H \phi)=(\psi',\phi)$ with $(.,.)$
the scalar product on the Hilbert space. By definition, we set
$H^\dag \psi=\psi'$.

\noindent An operator $H$ is called {\it symmetric} if for all
$\phi,\psi \in {\mathcal D}(H)$, we have $(H \phi,\psi)=(\psi,
H\psi)$ . $H$ is {\it self-adjoint} if additionally ${\mathcal
D}(H^\dag)={\mathcal D}(H)$. Let us assume that $(H,{\mathcal
D}(H))$ is densely defined, symmetric and closed with adjoint
$(H^\dag,{\mathcal D}(H^\dag))$. The {\it deficiency indices} are
then defined by

\bea n_\pm=Ker(H^\dag \mp i Id) \eea

\begin{thm}
For an operator with deficiency indices $(n_-,n_+)$, there are
three possibilities:

\begin{enumerate}
\item If $n_-=n_+=0$, then $H$ is self-adjoint (necessary and
sufficient condition)

\item If $n_+=n_-=n \geq 1$, then $H$ has infinitely many
self-adjoint extensions, parameterized by the unitary group
$U(n)$.

\item If $n_- \neq n_+$, then $H$ has no self-adjoint extension.

\end{enumerate}

\end{thm}

\noindent Note that if the deficiency indices are given by $(n,n)$
then the spectrum is discrete. Moreover, if $H$ admits a
self-adjoint extension, the resulting conditional probability is
not unique but depends on the boundary conditions which are
parameterized by the unitary  group $U(n)$.} (\cite{ree} and see
\cite{bon} for a pedagogical introduction).

\noindent After proving that $D$ is a self-adjoint operator or
admits a self-adjoint extension, the conditional probability (or
the fair value) can then be expanded over a complete basis of
orthonormal  eigenvectors $\phi_n(x)$

\bea p(x,\tau,x_0)=\sum_n e^{-E_n \tau} \phi_n(x) \phi_n(x_0) \eea

\noindent with $D\phi_n(x)=E_n\phi_n(x)$. The discrete summation
over $n$ can also include a continuous summation according to a
specific measure $\mu(E)$ if the spectrum contains a continuous
part.

\section{1D Time-homogeneous diffusion models}
In the next section, we apply the general reduction method,
presented previously, to the one-dimensional KBS equation. Similar
reduction to a Schrodinger equation with a scalar potential
(without any references to differential geometry) can be found in
\cite{car,lip,lin}. We then find the supersymmetric partner of
this Schrodinger equation and show how to generate new exact
solutions (for Vanilla options).

\subsection{Reduction method}

\noindent Consider a one-dimensional, time-homogeneous diffusion
process with drift \footnote{The time-dependent process
$df=\mu(f)A^2(t)dt+\sigma(f)A(t)dW$ is equivalent to this process
under the change of local time $t'=\int^t A^2(s)ds$}

\bea df=\mu(f)dt+\sigma(f)dW \eea

\noindent This process has been used as the basis for various
mathematical models in finance. If $f$ is a traded asset (i.e. a
forward), the drift vanishes in the forward measure and we have a
local volatility model where we assume that the volatility is only
a function of $f$.  The one-dimensional process is not necessarily
driftless as the random variable is not a traded asset as it is
the case for an instantaneous short rate model, or a model of
stochastic volatility.

\noindent In our framework, this process corresponds to a
(one-dimensional) real curve endowed with the metric $g_{ff}={2
\over \sigma(f)^2}$. For the new coordinate

\bea s(f)=\sqrt{2}\int_{f_0}^f {df' \over \sigma(f')} \eea

 \noindent the metric is
flat: $g_{ss}=1$. The Laplace-Beltrami operator therefore becomes
$\vartriangle=\partial^2_s$.

\noindent Using the definition (\ref{aaa}), (\ref{qqq}), we find
that the Abelian connection ${\mathcal A}_f$ and the function $Q$
are given by

\bea &&{\mathcal A}_f=-{1 \over 2} \partial_f{ln(\sigma(f))}+{\mu(f) \over \sigma^2(f)}   \label{ceva}\\
&&Q={1 \over 4}(\sigma(f) \sigma^{''}(f) -{1 \over 2}\sigma'(f)^2)-{
\mu'(f) \over 2}+{\mu(f) \sigma'(f) \over \sigma(f)}-{\mu(f) \over 2
\sigma^2(f)} \label{cevqq}\eea

\noindent In this case, by applying a gauge transformation on the
conditional probability $p$, $P={ \sigma(f_0)  \over \sqrt{2}}
e^\Lambda p$ with \bea \Lambda=-{1 \over 2} {ln({\sigma(f) \over
\sigma(f_0)})}+\int_{f_0}^f { \mu(f') \over \sigma^2(f')} df' \eea
\noindent then the connection vanishes and $P$ satisfies a heat
kernel with a scalar potential $Q(s)$ (in the $s$ flat coordinate)

\bea \partial_\tau P(s,\tau)= (\partial_s^2 +Q(s)) P(s,\tau)
\label{sch}\eea

\noindent The solution $P$ has been scaled by the (constant)
factor ${ \sigma(f_0)  \over \sqrt{2}}$ in order to have the
initial condition $lim_{\tau \rightarrow 0}P(s,\tau) =\delta(s)$.
Moreover, $Q$ is given in the $s$ coordinate by

\bea Q={1 \over 2}(ln(\sigma))^{''}(s)- {1 \over
4}((ln(\sigma))^{'}(s))^2-{ \mu'(s) \over \sqrt{2}
\sigma(s)}+{\sqrt{2} \mu(s) \sigma'(s) \over
\sigma(s)^2}-{\mu^2(s) \over 2 \sigma^2(s)} \label{QQ}\eea

\noindent where the prime $'$ indicates a derivative according to
$s$.

\begin{exa}[quadratic volatility process]
\noindent Let us assume that $f$ satisfies a driftless process
(i.e. $\mu(f)=0$). The Black-Scholes equation reduces to the heat
kernel on $\RR$ if $Q(s)=constant$ (i.e. $Q(s)$ is zero modulo a
time-dependent gauge transformation) which is equivalent to
$\sigma(f)=\alpha f^2 + \beta f+ \gamma$ (i.e. the quadratic
volatility model, also called the hyperbolic model \cite{lip}).
\end{exa}

\begin{exa}[CEV process]
For the CEV process $df= f^\beta dW_t$ $\mu(f)=0$, the potential
is  $Q(s)={\beta (\beta -2) \over 4(1-\beta)^2 s^2}$ for $\beta
\neq 1$ and $Q(s)=-{1 \over 8}$ for $\beta=1$.
\end{exa}

\noindent If the Hamiltonian $D=(-\partial_s^2 -Q)$ is
self-adjoint or admits self-adjoint extensions, the spectral
decomposition can be used and the conditional probability can be
decomposed using a complete basis of orthonormal eigenvectors
$\phi_n(s)$.

\bea P(s,\tau)=\sum_n \phi_n(s) \phi^\dag_n(s_0) e^{-E_n \tau}
\label{exp} \eea

\noindent with $\phi_n(s)$ satisfying

\bea  D\phi_n(s)=E_n \phi_n(s)   \label{eig}\eea

\vskip 3truemm

\noindent {\bf Boundary conditions}

\vskip 3truemm

\noindent  For a one-dimensional diffusion process, one don't need
to use the deficiency indices technique as the complete
classification of the boundary conditions is given by {\it
Feller's classification}. More precisely, for a 1D
time-homogeneous diffusion process, the boundary falls into one of
the four following types: regular, entrance, exit or natural. For
entrance, exit or natural, no boundary conditions are needed. For
a regular boundary, the conditional probability is not unique but
depends on the boundary conditions. It corresponds to the case
when the deficiency indices are $(n,n)$. The boundary
classification depends on the behavior of the following functions
$S(c,d)=\int_c^d s(f) df$, $M(c,d)=\int_c^d m(f) df$,
$\Sigma(e)=lim_{c \rightarrow e} \int_c^d S(c,f) m(f) df$ and $
N(e)=lim_{c \rightarrow e} \int_c^d S(x,d) m(x) dx$ with
$s(f)=e^{-2 \int^f {\mu(x) dx \over \sigma(x)^2}}$ and $m(f)={1
\over \sigma^2(f) s(f)}$ (see Table 1 below).

\begin{table}[h]
\begin{center}
\begin{tabular}{|c|cccc|}
\hline
Boundary type & $S(e,d)$ & $M(e,d)$ & $\Sigma(e)$ & $N(e)$ \\
\hline Regular & $< \infty$ & $< \infty$ & $< \infty$ & $< \infty$
\\
\hline Exit & $< \infty$ & $= \infty$ & $< \infty$ & $< \infty$
\\
\hline Entrance & $= \infty$ & $ < \infty$ & $< \infty$ & $<
\infty$
\\
\hline Natural & $< \infty$ & $= \infty$ & $= \infty$ & $= \infty$
\\
 & $= \infty$ & $< \infty$ & $= \infty$ & $= \infty$
\\
 & $= \infty$ & $= \infty$ & $= \infty$ & $= \infty$
\\
\hline
\end{tabular}
\caption{Feller's Classification.}
\end{center}
\end{table}

\vskip 3truemm

\begin{exa}[Vanilla option with constant interest rate]

\noindent We have that the forward $f$ satisfies a driftless
process (i.e. $\mu(f)=0$). The value at $t$ of a European option
(with strike $K$ and expiry date $T$) is then given by
($\tau=T-t$)

\bea {\mathcal C }(f_t,K,\tau)=e^{-r \tau} \int_K^\infty (f-K)
p(f,\tau|f_t) df \eea

\noindent Doing an integration by parts, or equivalently, applying
the Tanaka-Meyer formula on the payoff $(S_t-K)^+$ \cite{alb2} we
can show that the vanilla option ${\mathcal C}$ can be rewritten
as

\bea {\mathcal C }(f_t,K,\tau)=e^{-r \tau}(f_t-K)^++e^{-r
\tau}{\sigma^2(K) \over 2}\int_0^\tau dt' p(K,t'|f_t) \eea

\noindent Using the relation between the conditional probability
$p(f,t'|f_t)$ and its gauge-transform $P(s(f),t'|s_t)$, we obtain

\bea {\mathcal C }(f_t,K,\tau)=e^{-r \tau}((f_t-K)^++{\sigma(K)^{5
\over 2} \over \sqrt{2} \sigma(f_t)^{3 \over 2} } \int_0^\tau
P(s(K),t'|s_t) dt') \label{opt1} \eea

\noindent Plugging the expression for $P(s,t'|s_t)$ (\ref{exp})
into (\ref{opt1}) and doing the integration over the time $t$, we
obtain \cite{alb2}

\bea {\mathcal C }(f_t,K,\tau)=e^{-r \tau}((f_t-K)^++{\sigma(K)^{5
\over 2} \over \sqrt{2} \sigma(f_t)^{3 \over 2} } \sum_n
\phi_n(s(K)) \phi_n(s_t) {(1-e^{-E_n \tau})  \over E_n})
\label{option} \eea

\end{exa}

\noindent A specific local volatility model will give an exact
solution for a vanilla option if we can find the eigenvalues and
eigenvectors for the corresponding Euclidean Schrodinger equation
with a scalar potential. As examples of solvable potentials, one
can cite the harmonic oscillator, Coulomb, Morse,  Poschl-Teller
I\&II, Eckart and Manning-Rosen potentials. The classification of
solvable scalar potentials was initiated by Natanzon \cite{nat}.
This work provides the most general potential for which  the
Schrodinger equation can be reduced to either a hypergeometric or
confluent equation. We will review in the following section the
{\it Natanzon potential}, which depends on $6$ parameters. We will
later show that the Schrodinger equation can be doubled into a set
of two independent Schrodinger equations with two different scalar
potentials  which transform into each other under a supersymmetric
transformation. Moreover, if one scalar potential is solvable, the
other one is. Applying this technique to the Natanzon potential,
we will obtain a new class of solvable potentials corresponding to
a new class of solvable diffusion processes. For these models, the
solution to the KBS equation is given  by a sum of hypergeometric
functions.

\subsection{Solvable (super)potentials}

In this section, we show that the Schrodinger equation can be
formulated using supersymmetric techniques (see \cite{coo} for a
nice review). In particular, the local volatility will be
identified as a superpotential. Using this formalism, we show how
to generate a hierarchy of solvable diffusion models starting from
a known solvable superpotential, for example the hypergeometric or
confluent hypergeometric Natanzon superpotential.

\subsubsection{Superpotential and local volatility}

\noindent Let us write the Kolmogorov equation (\ref{eig}) in the
following way by introducing the first-order operator $A_1$ and
its  formal adjoint $A^{\dag}_1$ \footnote{In order to obtain the
correct adjoint operator on $\RR^+$, we impose the absorbing
boundary condition $\phi_n(s=0)=0$}

 \bea  E^{(1)}_n
{\phi}^{(1)}_n=  A^\dag_1 A_1 {\phi}^{(1)}_n \label{susy}\eea

\noindent with $A_1=\partial_s + W^{(1)}(s) \;, \;
A^\dag_1=-\partial_s + W^{(1)}(s)$. $W^{(1)}$ is called the {\it
superpotential} which satisfies the Riccatti equation

\bea Q^{(1)}(s)=\partial_s W^{(1)}(s)-{W^{(1)}}^2(s) \eea

\noindent Surprisingly, this equation is trivially solved for our
specific expression for $Q^{(1)}$ (\ref{QQ}) (even with a drift
$\mu(f)$!)

\bea W^{(1)}(s)={1 \over 2}{ d ln\sigma^{(1)}(s) \over ds}
-{\mu^{(1)}(s) \over \sqrt{2} \sigma^{(1)}(s)} \label{superpot}
\eea

\noindent In particular, for zero drift, the local volatility
function is directly related to the superpotential by
$\sigma(s)=e^{2 \int^s W(z) dz}$. A similar correspondence between
the superpotential and driftless diffusion process has been found
in \cite{kry,jun}. Moreover, if we have a family of solvable
superpotentials $W_{solvable}^{(1)}(s)$, we can always find an
analytic solution to the Kolmogorov equation for any diffusion
term $\sigma^{(1)}(s)$ by adjusting the drift with the relation
(\ref{superpot})

\bea \mu^{(1)}(s) ={\sigma^{(1)'}(s) \over
\sqrt{2}}-\sqrt{2}\sigma^{(1)}(s)W_{solvable}^{(1)}(s) \eea

\noindent Note that (\ref{susy}) admits a zero eigenvalue if and
only if the Kolmogorov equation admits a stationary distribution.
By observing that $A^\dag_1 A_1 \phi^{(1)}_0=0$ is equivalent to
$A_1 \phi^{(1)}_0=0$, we obtain the stationary distribution

\bea \phi^{(1)}_0(s)=C e^{-\int^s W^{(1)}(z)dz} \label{ground}
\eea

\noindent with $C$ a normalization constant. Therefore, the
stationary distribution will exist if the superpotential is
normalisable (i.e. $\phi^{(1)}_0(s) \in L^2$).

\begin{exa}[Coulomb superpotential and CEV process]
The CEV process corresponding to $\sigma(f)=\sigma_0 f^\beta$ and
$\mu(f)=0$ has the {\it Coulomb superpotential} $W(s)={\beta \over
2(1-\beta)s}$.
\end{exa}

\noindent Next, we define the {\it Scholes-Black equation} by
intervening the operator $A_1$ and $A^\dag_1$

\bea E^{(2)}_n  {\phi}^{(2)}_n(s)&=&A_1 A^\dag_1  {\phi}^{(2)}_n(s) \\
&=&(-\partial^2_s-Q^{(2)}(s)){\phi}^{(2)}_n(s) \label{susy1}\eea

\noindent We obtain a new Schrodinger equation with the partner
potential \bea Q^{(2)}(s)=-\partial_sW^{(1)}-(W^{(1)})^2
\label{partner}\eea

\noindent Plugging our expression for the superpotential
(\ref{superpot}) in (\ref{partner}), we have

\bea {Q}^{(2)} =-{1 \over 2}(ln(\sigma^{(1)}))^{''}(s)- {1 \over
4}(ln(\sigma^{(1)}))^{'}(s))^2+{ {\mu^{(1)}}'(s) \over \sqrt{2}
\sigma^{(1)}(s)}-{{\mu^{(1)}}^2(s) \over 2 {\sigma^{(1)}}^2(s)}
 \label{partnerQ} \eea

\noindent In the same way as before, $H_2$ admits a zero
eigenvector (i.e. stationary distribution) if $\phi^{(2)}_0(s)=C
e^{\int^s W^{(1)}(z)dz}$ is normalisable.

\begin{rem}
In physics, the supersymmetry (SUSY) is said to be broken if at
least one of the eigenvectors $\phi^{(1,2)}_0(s)$ exists.
Otherwise, SUSY is said to be broken dynamically.
\end{rem}

\noindent Now we want to show that if we can solve the equation
(\ref{susy}) then we have automatically a solution to
(\ref{susy1}) and vice-versa. The SUSY-partner Hamiltonians
$H_1=A^\dag_1 A_1$ and $H_2=A_1 A^\dag_1 $ obey the relations
$A_1^\dag H_2=H_1 A_1^\dag$ and $H_2 A_1=A_1 H_{1}$. As a
consequence $H_1$ and $H_2$ are isospectral. More precisely, the
strictly positive eigenvalues all coincide and the corresponding
eigenvectors are related by the supercharge operators $A_1$ and
$A^\dag_1$:

\vskip 3truemm

\noindent $\rhd$ If $H_1$ admits a zero eigenvalue (i.e. broken
supersymmetry), we have the relation

\bea &&E_n^{(2)}=E_{n+1}^{(1)} \; ; \ E_{0}^{(1)}=0 ; \;  \phi^{(1)}_0(s)=C e^{-\int^s W^{(1)}(z)dz}  \\
&&\phi^{(2)}_n(s)=(E_{n+1}^{(1)})^{-{1 \over 2}} A_1 \phi^{(1)}_{n+1}(s) \\
&&\phi^{(1)}_n(s)=(E_n^2)^{-{1 \over 2}} A^\dag_1 \phi^{(2)}_n(s)
\label{susyt1} \eea \noindent

\noindent $\rhd$ If $H_1$ (and $H_2$) doesn't admit a zero
eigenvalue (i.e. unbroken supersymmetry)

\bea &&E_n^{(2)}=E_{n}^{(1)}\\
&&\phi^{(2)}_n(s)=(E_{n}^{(1)})^{-{1 \over 2}} A_1 \phi^{(1)}_{n}(s) \\
&&\phi^{(1)}_n(s)=(E_n^2)^{-{1 \over 2}} A^\dag_1 \phi^{(2)}_n(s)
\eea

\noindent In the broken SUSY case, there are no zero modes and
consequently the spectrum of $H_1$ and $H_2$ are the same. One can
then obtain the solution to  the Scholes-Black (resp.
Black-Scholes) equation if the eigenvalues/eigenvectors of the
Black-Scholes (resp. Scholes-Black) are known. We  clarify this
correspondence by studying a specific example: the CEV process
$df=f^\beta dW$. In particular, we show that  for $\beta={2 \over
3}$, the partner superpotential vanishes. It is therefore simpler
to solve the Scholes-Black equation as Scholes-Black (rather than
Black-Scholes) reduces to the heat kernel on $\RR^+$. Applying a
supersymmetric transformation on the Scholes-Black equation, we
can then derive the solution to the Black-Scholes equation.

\begin{exa}[CEV with $\beta={2/3}$ and Bachelier process]
We saw previously that the superpotential associated with the CEV
process is given by

\bea W^{(1)}(s)={\beta \over 2 s(1-\beta)} \eea

\noindent with the flat coordinate  $s= {\sqrt{2} f^{1-\beta}
\over (1-\beta)}$ and the potential (\ref{QQ})

\bea Q^{(1)}(s)={\beta(\beta-2) \over 4(1-\beta)^2 s^2} \eea

\noindent from which we deduce the partner potential
(\ref{partnerQ})

\bea Q^{(2)}(s)={\beta (2-3\beta) \over 4(1-\beta)^2 s^2} \eea

\noindent This partner potential corresponds to the potential of a
CEV process $df=f^B dW$ with $B$ given as a function of $\beta$ by

\bea {B(B-2) \over (1-B)^2}={\beta (2-3\beta) \over (1-\beta)^2}
\eea

\noindent Surprisingly, we observe that for $\beta={2 \over 3}$,
$Q^{(2)}(s)$ cancels and the corresponding partner local
volatility model is the Bachelier model $df=dW$ for which the heat
kernel is given by the normal distribution. The eigenvectors of
the supersymmetric Hamiltonian partner $H_2=-\partial_s^2$ to
$H_1$ are given by (with the absorbing boundary condition
$\phi_n(0)=0$)

\bea \phi^{(2)}(s,E)={ sin({\sqrt{E}s}) \over \sqrt {4 \pi} E^{1
\over 4}} \eea

\noindent with a continuous spectrum $\RR^+$. Applying the
supersymmetric transformation (\ref{susyt1}), we obtain the
eigenvectors for the Hamiltonian $H_1=-\partial_s^2+{2 \over s^2}$
corresponding to the CEV process with $\beta={2 \over 3}$

\bea \phi^{(1)}(s,E)&=&E^{-{1 \over 2}}(-\partial_s + { 1 \over
s})
\phi^{(2)}(s,E) \\
&=&{ 1 \over \sqrt {4 \pi} E^{3 \over
4}}(-\sqrt{E}cos({\sqrt{E}s})+{sin({\sqrt{E}s}) \over s}) \eea

\noindent Plugging this expression in (\ref{option}), we obtain
the fair value for a vanilla option which can be integrated and
written in terms of the cumulative distribution \cite{lip}

\beaa &&e^{r\tau}{\mathcal C}(f_t,K,\tau)=(f_t-K)^++{ K^{ 5 \over
3} \over \sqrt{2} f_t} \int_0^\infty dE  { (1-e^{-E \tau}) \over
E^{}}\phi^{(2)}(s,E)\phi^{(2)}(s_0,E) \eeaa

\noindent The fact that the CEV model with $\beta={2 \over 3}$
depends on the normal cumulative distribution and is therefore
related to the heat kernel on $\RR^+$ has been observed
empirically by \cite{lip}. Here we have seen that it corresponds
to the fact that the supersymmetric partner potential vanishes for
this particular value of $\beta$.

\end{exa}

\subsection{Hierarchy of solvable diffusion processes}

\noindent In the previous section we saw that the operators $A_1$
and $A_1^{\dag}$ can be used to factorize the Hamiltonian $H_1$.
These operators depend on the superpotential $W^{(1)}$ which is
determined once we know the first eigenvector $\phi^{(1)}_0(s)$ of
$H_1$ (\ref{ground}). We have assumed that $H_1$ admits a zero
eigenvalue. By shifting the energy $E$ it is always possible to
achieve this condition. The partner Schrodinger equation
(\ref{susy1}) can then be recast into a Schrodinger equation with
a zero eigenvalue

\bea H_{(2)}=A_1 A_1^\dag = A^\dag_2 A_2+E_1^{(1)} \eea

\noindent where $A_2 \equiv \partial_s+W_2(s)$ and $A_2^\dag
\equiv -\partial_s+W_2(s)$,

\bea W^{(2)}(s) \equiv {1 \over 2}{ d ln\sigma^{(2)}(s) \over ds}
-{\mu^{(2)}(s) \over \sqrt{2} \sigma^{(2)}(s)} \label{super2} \eea

\noindent We have introduced the notation that in $E_{n}^{(m)}$,
$n$ denotes the energy level and $(m)$ refers to the
$\mathrm{m^{th}}$ Hamiltonian $H_m$. By construction, this new
Hamiltonian $H_2= A^\dag_2 A_2+E_1^{(1)}$ is solvable as $A_1
A_1^\dag$ is and the associated diffusion process with volatility
$\sigma^{(2)}$ and drift $\mu^{(2)}$ satisfy (\ref{super2}). The
superpotential $W^{(2)}(s)$  is determined by the first
eigenvector of $H_2$, $\phi^{(2)}_0(s)$,

\bea W^{(2)}(s)=-{d ln(\phi^{(2)}_0)(s) \over ds} \eea

\noindent We can then apply a supersymmetric transformation on
$H_2$. The new Hamiltonian $H_3$ can be refactorised exactly in
the same way we did for $H_2$. Finally, it is not difficult to see
that if $H_1$ admits $p$ discrete (normalisable) eigenvectors,
then one can generate a family of solvable Hamiltonians $H_m$
(with a zero-eigenvalue by construction)

\bea H_m=A_m^\dag A_m + E_{m-1}^{(1)}=-\partial_s^2+Q_m(s) \eea

\noindent where $A_m=\partial_s+W_m(s)$. This corresponds to the
solvable diffusion process with a drift and a volatility such that

\bea W_m(s)=-{d ln \phi^{(m)}_0 \over ds}={1 \over 2}{ d
ln\sigma^{(m)}(s) \over ds} -{\mu^{(m)}(s) \over \sqrt{2}
\sigma^{(m)}(s)} \eea

\noindent The eigenvalues/eigenvectors of $H_m$ are related to
those of $H_1$ by

\bea &&E_n^{(m)}=E_{n+1}^{(m-1)}=\cdots =E_{n+m-1}^{(1)} \\
&&\phi_n^{(m)}=(E^{(1)}_{n+m-1}-E^{(1)}_{m-2})^{-{1 \over 2}}
\cdots (E^{(1)}_{n+m-1}-E^{(1)}_{0})^{-{1 \over 2}}A_{m-1} \cdots
A_1 \phi_{n+m-1}^{(1)} \eea

\noindent In particular, the superpotential of $H_m$ is determined
by the $(m-1)^{\mathrm th}$ eigenvector of $H_1$,
$\phi^{(1)}_{m-1}(s)$,

\bea W_m(s)&=&-{d ln (A_{m-1} \cdots A_1 \phi^{(1)}_{m-1}(s))
\over
ds}\\
&=&{1 \over 2}{ d ln\sigma^{(m)}(s) \over ds} -{\mu^{(m)}(s) \over
\sqrt{2} \sigma^{(m)}(s)} \label{res1} \eea

\noindent Consequently, if we know all the $m$ discrete
eigenvalues and eigenvectors of $H_1$, we immediately know all the
energy eigenvalues and eigenfunctions of the hierarchy of $m-1$
Hamiltonians. In the following we  apply this procedure starting
from a known solvable superpotential, {\it the Natanzon}
superpotential.

\subsection{Natanzon (super)potentials}

\noindent The Natanzon potential \cite{nat} is the most general
potential which allows us to reduce the Schrodinger equation
(\ref{sch}) to a Gauss or confluent hypergeometric equation (GHE
or CHE).

\subsubsection{Gauss hypergeometric potential}

The potential is given by

\bea Q(s)={S(z)-1 \over R(z)}-({ r_1-2(r_2+r_1)z \over z(1-z)}-{5
\over 4}{(r_1^2-4 r_1 r_0) \over R(z)}+r_2){ z^2(1-z^2) \over
R(z)^2} \eea

\noindent with  $R(z)=r_2 z^2+r_1 z+r_0 \; >0$ and $S(z)=s_2
z^2+s_1 z+s_0$ (two second order polynomials). The $z$ coordinate,
lying in the interval $[0,1]$, is defined implicitly in terms of
$s$ by the differential equation

\bea {dz(s) \over ds}={2 z(1-z) \over \sqrt{R(z)} } \eea

\begin{exa}
\noindent The hypergeometric Natanzon potential includes as
special cases the Posch-Teller potential II $Q(s)=A+B sech({2 s
\over \sqrt{r_1}})^2+C csch({2 s \over \sqrt{r_1}})^2$ for
$r_0=r_2=0$ and the Rosen-Morse potential $Q(s)=A+B tanh({2 s
\over \sqrt{r_0}})+C sech({2 s \over \sqrt{r_0}})^2$, $r_1=r_2=0$
with $A$ and $B$ constant.
\end{exa}

\noindent By construction, the solution to the Schrodinger
equation with a GHE potential is given in terms of the Gauss
hypergeometric function $F(\alpha,\beta,\gamma,z)$

\bea  \psi(s,E)=(z')^{-{1 \over 2}} z^{\gamma \over
2}(1-z)^{{-\gamma+\alpha+\beta+1 \over 2}}
F(\alpha,\beta,\gamma,z)
  \eea

\noindent where $F(\alpha,\beta,\gamma,z)$ satisfies the
differential equation \cite{abr}

 \bea z(1-z){d^2 F \over dz^2}+(\gamma-(\alpha+\beta+1)z){d F \over
 dz}-\alpha \beta z=0 \label{hyper1} \eea
\noindent and  the most general solution to this equation
(\ref{hyper1}) is generated by a two-dimensional vector space

\bea F(\alpha,\beta,\gamma,z)=c_1
{{}_2}\,{F_1}(\alpha,\beta,\gamma,z)+ c_2 z^{1-\gamma}
{{}_2}\,{F_1}(\alpha-\gamma+1,\beta-\gamma+1,2-\gamma,z) \eea

\noindent with ${{}_2}\,{F_1}(\alpha,\beta,\gamma,z)$ satisfying
${{}_2}\,{F_1}(\alpha,\beta,\gamma,0)=1$ and $c_1$ and $c_2$ two
arbitrary coefficients. The parameters $\alpha$, $\beta$, $\gamma$
depend explicitly on the eigenvalue $E$  by

\bea 1-(\alpha-\beta)^2= r_2 E+s_2 \\
2 \gamma (\alpha+\beta-1)-4 \alpha \beta=r_1 E+s_1 \\
\gamma(2-\gamma)=r_0 E+s_0 \eea

\noindent In fact, one can show that $E$ satisfies a fourth-order
polynomial and find  $E$ as an explicit function of $\alpha$,
$\beta$, $\gamma$ \cite{gro}. By imposing the condition that the
eigenvectors are normalisable (i.e. belong to $L^2([0,1])$) we
obtain the discrete spectrum $E_n$ and can determine the
coefficients $c_1$ and $c_2$. We impose conditions on $\alpha$,
$\beta$, $\gamma$, $c_1$ and $c_2$ such that

\bea \int_0^1 dz {R(z) \over 4} z^{\gamma -2}
(1-z)^{-\gamma+\alpha+\beta-1} F^2(\alpha,\beta,\gamma,z) < \infty
\eea

\noindent Looking at the asymptotic behavior of
${{}_2}{F_1}(\alpha,\beta,\gamma,z)$ near $z=0$ and $z=1$
\cite{abr} \footnote{ ${}_2F_1(\alpha,\beta,\gamma,z)\sim_{z
\rightarrow 1}{\Gamma}(\gamma )\left( {\left( -1 + {z} \right) }^
      {-{\alpha } - \beta  + {\gamma }}
     + \frac{{\Gamma}(-\alpha  - \beta  + \gamma )}
      {{\Gamma}(-\alpha  + \gamma ){\Gamma}(-\beta  + \gamma )}
      \right)$}, we
obtain the following conditions

\begin{table}[h]
\begin{center}
\begin{tabular}{|c|c|}
\hline $r_0 \neq 0$ & $c_1=0, \gamma <1$ or $c_2=0 , \gamma >1$
\\\hline
$r_0=0$, $r_1 \neq 0$ & $c_1=0, \gamma <2$ or $c_2=0 , \gamma >0$
\\ \hline
$r_0=0$, $r_1 = 0$ & $c_1=0, \gamma <3$ or $c_2=0 , \gamma >-1$ \\
\hline
\end{tabular}
\caption{condition $z=0$} \label{figure0}
\end{center}
\end{table}

\begin{table}[h]
\begin{center}
\begin{tabular}{|c|c|}
\hline $c_1=0$ & $\alpha-1 \in \NN^*$, $\alpha+\beta-\gamma <0$ or
$-1-\alpha+\gamma \in \NN^*$ , $\alpha+\beta-\gamma >0$ \\ \hline
$c_2=0$ & $-\alpha \in \NN^*$, $\alpha+\beta-\gamma >0$ or $\alpha-\gamma \in \NN^*$ , $\alpha+\beta-\gamma <0$ \\
\hline
\end{tabular}
\caption{condition $z=1$} \label{figure1}
\end{center}
\end{table}

\noindent We have the discrete eigenvalues ($\alpha_n=-n \; ; n
\in \NN^*$)

\bea &&2n+1=-\sqrt{1-r_0E_n-s_0}+\sqrt{1-r_2
E_n-s_2}-\sqrt{1-(r_0+r_1+r_2)E_n-(s_0+s_1+s_2)} \\
&&\psi_n(s)=(z')^{-{1 \over 2}} z^{\gamma_n \over
2}(z-1)^{{-\gamma_n+\alpha_n+\beta_n+1 \over 2}}
F(\alpha_n,\beta_n,\gamma_n,z) \\
&&\gamma_n=1 + \sqrt{1-r_0E_n-s_0} \\
&&\alpha_n-\beta_n=- \sqrt{1-r_2E_n-s_2}\\
&&\alpha_n+\beta_n=\gamma_n+\sqrt{1-(r_0+r_1+r_2)E_n-(s_0+s_1+s_2)}\eea

\noindent and the (normalised) eigenvectors

\bea &&\psi_n(s)=B_n (z')^{-{1 \over 2}} z^{\gamma_n \over
2}(1-z)^{{-\gamma_n-n+\beta_n+1 \over 2}}
P_n^{(\gamma_n-1,-n+\beta_n-\gamma_n+1)}(1-2z) \eea

\noindent with $B_n=[(({R(1) \over \alpha+\beta-\gamma})+({r_0
\over \gamma-1})-({r_2 \over \beta-\alpha})){ \Gamma(\gamma+n+1)
\Gamma(\alpha+\beta-\gamma) \over n!\Gamma(\beta-\alpha-n)}]^{-{1
\over 2}}$ and $P_n^{(\gamma-1,\alpha+\beta-\gamma)}(2z-1)$ the
Jacobi polynomials.

\subsubsection{Confluent hypergeometric potential}

A similar construction can be achieved for the class of {\it
scaled confluent hypergeometric functions}. The potential is given
by

\bea Q(s)={S(z)-1 \over R(z)}-({r_1 \over z}-{5 \over
4}{(r_1^2-4r_2 r_0) \over R(z)}-r_2){z^2 \over R(z)^2} \eea

\noindent with  $R(z)=r_2 z^2+r_1 z+r_0 \; >0$ and $S(z)=s_2
z^2+s_1 z+s_0$. The $z$ coordinate, lying in the interval
$[0,\infty[$, is defined implicitly in terms of $s$ by the
differential equation \bea {dz(s) \over ds}={2 z \over
\sqrt{R(z)}} \eea

\begin{exa}
\noindent The confluent Natanzon potential reduces to the Morse
potential $Q(s)={-1+s_0+s_1 e^{2 s \over \sqrt{r_0}}+s_2  e^{4 s
\over \sqrt{r_0}} \over r_0}$ for $r_1=r_2=0$, to the 3D
oscillator $Q(s)={ -{3 \over 4}+s_0 \over s^2}+{s_1 \over r_1}
+{s_2 s^2 \over r_1^2}$ for $r_0=r_2=0$ and to the Coulomb
potential $Q(r)={-r_2 s_0-2 s \sqrt{r_2} s_1 -4 s^2 s_2 \over 4
r_2 s^2}$ for $r_0=r_1=0$.
\end{exa}

\noindent By construction, the solution to the CHE potential is
given in terms of the confluent hypergeometric function
$F(\alpha,\beta,\gamma,z)$

\bea \psi(s,E)=z(s)^{\gamma \over 2} e^{-{\omega z(s) \over 2}}
(z'(s))^{-{1 \over 2}} F(\alpha,\beta,\gamma,\omega z(s))\eea

\noindent The parameters $\omega$, $\gamma$, $\alpha$ depend
explicitly on the eigenvalue $E$  by

\bea &&\omega^2=-r_2 E-s_2 \\
&&2 \omega (\gamma-2\alpha)=r_1 E+s_1 \\
&&\gamma(2-\gamma)=r_0 E+s_0 \eea

\noindent Note that $\phi(z) \doteq F(\alpha,\beta,\gamma,\omega
z(s))$ satisfies the differential equation \cite{abr}

\bea z \phi''(z)+(\gamma-\omega z)\phi'(z)-\omega \alpha \phi(z)=0
\label{hyper2} \eea
 \noindent and the most general solution to
(\ref{hyper2}) is generated by a two-dimensional vector space \bea
F(\alpha,\gamma,\omega z)=c_1 M(\alpha,\gamma,\omega z)+c_2
(\omega z)^{1-\gamma} M(1+\alpha-\gamma,2-\gamma,\omega z) \eea
\noindent with $M(\alpha,\gamma,\omega z)$ the M-Whittaker
function and $c_1$ and $c_2$ two arbitrary coefficients. By
imposing that the eigenvectors are normalisable ((i.e. belong to
$L^2(\RR^+)$)) , we obtain the following conditions (see Table 2
\&5) which give the spectrum $E$ and the coefficients $c_1$ and
$c_2$ \footnote{ $M(\alpha,\beta,z) \sim_{z\rightarrow\infty}
{\Gamma(\beta) \over \Gamma(\alpha)}e^{{z}}
z^{\alpha-\beta}(1+O(|z|^{-1}))$}

\begin{table}[h]
\begin{center}
\begin{tabular}{|c|c|}
\hline $\alpha>2$ & no condition
\\
\hline $\alpha \leq 2$ & $c_1=0, -1-\alpha+\gamma \in \NN^*$
\\
 & or $c_2=0 , -\alpha \in \NN^*$ \\
 \hline
\end{tabular}
\caption{condition $z = \infty$} \label{fig2}
\end{center}
\end{table}

\noindent We have the discrete eigenvalues $\alpha_n=-n \; ; n \in
\NN$

\bea\gamma_n&=&1 + \sqrt{1-r_0 E_n-s_0} \\
\omega_n&=&\sqrt{-r_2 E_n -s_2} \\
2n+1&=&{r_1 E_n +s_1 \over 2 \sqrt{-r_2 E_n-s_2}} -\sqrt{1-r_0 E_n
-s_0} \eea

\noindent and the (normalized) eigenvectors

\bea \psi_n(s)={n! \over (\gamma_n)_n} z(s)^{\gamma_n \over 2}
e^{-{\omega_n z(s) \over 2}} (z'(s))^{-{1 \over 2}}
L_n^{\gamma_n-1}(\omega_n z(s)) \eea

\noindent with $L_n^{\gamma_n-1}(z)$ the (generalized) Laguerre
polynomial. In the following, we have listed classical solvable
superpotentials and the corresponding solvable local volatility
models and solvable instantaneous short-rate models (Table 5 \&
6).

\begin{table}[h]
\begin{center}
\begin{tabular}{|c|c|c|}
\hline
Superpotential & $W(s)$ & ${\sigma(s) \over \sigma_0}$ \\
\hline Shifted oscillator & $a s+b$ & $ e^{a s^2+ 2 bs}$
\\
\hline Coulomb & $a +{b \over s}$ & $ s^b e^{2 as  }$
\\ \hline Morse & $a+b e^{-\alpha s}$ & $e^{2\left( -\left(
\frac{b}{\alpha e^{\alpha s}} \right)  + as \right) }$   \\
\hline Eckart& $acoth(\alpha s)+b$& $e^{2\left( bs + \frac{a\log
(\sinh (\alpha s))}{\alpha} \right) }$  \\
\hline  Rosen-Morse  & $a th(\alpha s) +b$& $e^{2 b s + {2a \over
\alpha}  ln(cosh(\alpha))}$   \\
\hline 3D oscillator & $as+{b \over s}$ & $e^{a s^2 +2 b ln(s)}$  \\
\hline P-T I $\alpha > 2b$& $a tan(\alpha s)+ b cotg(\alpha s)$ &
$e^{2\left( -\left( \frac{a\log (\cos (\alpha s))}{\alpha} \right)
+ \frac{b\log (\sin (\alpha s))}{\alpha} \right) }$
 \\
\hline P-T II $\alpha > 2b$& $a th(\alpha s)+ b coth(\alpha s)$ &
$e^{2\left( \frac{a\log (\cosh (\alpha s))}{\alpha} +
\frac{b\log (\sinh (\alpha s))}{\alpha} \right) }$ \\
\hline
\end{tabular}
\caption{Example of solvable superpotentials-Local Volatility}
\label{fig1}
\end{center}
\end{table}

\begin{table}[h]
\begin{center}
\begin{tabular}{|c|c|c|}
\hline one-factor short-rate model  & SDE & Superpotential \\
\hline Vasicek-Hull-White & $dr=k(\theta-r)dt+\sigma dW$ & Shifted
Osc. ($a={ \kappa \over 2}$, $b=-{\kappa \theta \over \sqrt{2} \sigma}$)\\
\hline
CIR & $dr=k(\theta-r)dt+\sigma \sqrt{r}dW$ & 3D Osc. ($a={\kappa \over 4}$, $b={1 \over 2}-{2 \theta \kappa \over \sigma^2}$) \\
\hline Doleans & $dr=k r dt+\sigma r dW$ & (Constant
 ($W(s)={-2 \kappa +\sigma^2 \over 2 \sqrt{2}
\sigma}$)
\\ \hline EV-BK & $dr=r(\eta-\alpha ln(r))dt+\sigma r dW$ & Shifted
Osc. ($a={\alpha \over 2}$, $b={\sqrt{2} (-2 \eta+\sigma^2) \over 4 \sigma}$)\\
\hline
\end{tabular}
\caption{Example of solvable one-factor short-rate models}
\label{fig1}
\end{center}
\end{table}

\subsubsection{Natanzon hierachy and new solvable processes}

\noindent We know that the Natanzon superpotential is related to
the zero-eigenvector

\bea W_{nat}=-\partial_s ln(\psi_0(s)) \eea

\noindent and the corresponding supercharge $A$ is

 \beaa &&A={\partial_s}+W_{nat}(s)
\\ &&={2 z(1-z) \over \sqrt{R}}(\partial_z -{\gamma_0 \over 2
z}-{(1+\alpha_0+\beta_0-\gamma_0) \over 2(z-1)}-{ \alpha_0 \beta_0
F(1+\alpha_0, 1+\beta_0,1+\gamma_0,z)  \over \gamma_0 F(\alpha_0,
\beta_0,\gamma_0,z)}+{ z'^{-{3 \over 2}} z''(s) \over 2 }) \eeaa

\noindent With a zero drift, the Natanzon superpotential
corresponds to the diffusion process (\ref{superpot})

\bea \sigma^{(1)}(s)={\sigma_0^{(1)} \over \phi_0^{(1)}(s)^{2}}
\eea

\noindent with $\sigma_0^{(1)}$ a constant of integration.
Applying the results of section (3.2), we obtain that the
driftless diffusion processes

\bea \sigma^{(m)}(s)={\sigma_0^{(m)} \over A_{m-1} \cdots A_1
\phi_{m-1}^{(1)}(s)^{2}} \eea

\noindent are solvable (\ref{res1}). Using the fact that the
(discrete) eigenvector $\phi_{m-1}^{(1)}(s)$ is a hypergeometric
function and that the derivative of a hypergeometric function is a
new hypergeometric function \footnote{ ${}_2
F_1'(\alpha,\beta,\gamma,z)={\alpha \beta \over \gamma} {}_2
F_1(\alpha+1,\beta+1,\gamma+1,z)$ and $M'(\alpha,\beta,z)={\alpha
\over \beta}M(\alpha+1,\beta+1,z)$} the action of $A_{m-1} \cdots
A_1$ on $\phi_{m_1}^{(1)}(s)$ will result in a sum of $(m-1)$
hypergeometric functions, thus generalizing the solution found in
\cite{alb}.

\section{Gauge-free stochastic volatility models}

\noindent In this section, we  try to identify the class of
time-homogeneous stochastic volatility models which leads to an
exact solution to the KBS equation. We assume that the forward $f$
and the volatility $a$ are driven by two correlated Brownian
motions in the risk-neutral measure

\bea &&df_t=a_t C(f_t) dW_1 \\
&&da_t=b(a_t,f_t)dt+\sigma(a_t,f_t) dW_2 \\
&&dW_1 dW_2=\rho dt \eea

\noindent with initial condition $a_0=a$ and $f_0=f$.

\noindent Using the definition for the connection (\ref{aaa}), we
obtain the Abelian connection \footnote{ \bea &&{\mathcal A}^f=-{
a^2 C \sigma \over 4}
\partial_f { C \over
\sigma} \\
&&{\mathcal A}^a={1 \over 2}(b-{a \sigma \over 2} \partial_a {
\sigma \over a}) \eea}

\bea &&{\mathcal A}_f=-{1 \over 2(1-\rho^2)}
\partial_f ln({C \over \sigma}) -{\rho \over (1-\rho^2)} ( { b \over a C \sigma}
-{1 \over 2C}\partial_a
{\sigma \over a} )\\
&& {\mathcal A}_a={1 \over (1-\rho^2)}( {b \over \sigma^2} -{1
\over 2}
\partial_a ln({ \sigma \over a}) )\eea

\noindent Then, the field strength is
\bea {\mathcal F}_{af}&=&\partial_a {\mathcal A}_f -\partial_f {\mathcal A}_a \\
&=& {1 \over (1-\rho^2)}[ (\partial_{af} ln(\sigma)-\partial_f {b
\over \sigma^2})-\rho({1 \over C}\partial_a { b \over a \sigma}-{1
\over 2 C}\partial^2_a { \sigma \over a}+{a \over 2}\partial_f^2 {
C \over \sigma}) ]\eea

\noindent We will now assume that the connection is flat,
${\mathcal F}_{af}=0$, meaning that the connection can be
eliminated modulo a gauge transformation. In this case, the
stochastic volatility model can thus be called a {\it gauge-free
model}. This condition is satisfied for every correlation $\rho$
if and only if

\bea && \partial_{af} ln(\sigma)=\partial_f {b \over
\sigma^2} \\
&& \partial_a { b \over a \sigma}-{1 \over 2 }\partial^2_a {
\sigma \over a}+{a C\over 2}\partial_f^2 { C \over \sigma}=0 \eea

\noindent Moreover, if we assume that $\sigma^a(a)$ is only a
function of $a$ (this hypothesis is equivalent to assuming that
the metric admits a Killing vector), the model is gauge-free if
and only if  \bea { b \over \sigma}={a \over 2}\partial_a {\sigma
\over a}+a \phi(f)-{a C(f) \over 2}
\partial_f^2 C(f) \int {a' da' \over \sigma(a')} \eea

\noindent with $\phi(f)$ satisfying \bea \partial_f
\phi(f)={\partial_f (C
\partial_f^2 C) \over 2}\int {a' da' \over \sigma(a')} \eea

\noindent This last equation is equivalent to $C(f) \partial_f^2
C(f) =\beta$ with $\beta$ a constant and $\phi=\gamma$ a constant
function. For $\beta=0$, the above equations simplify and we have

\bea &&C(f)=\mu f + \nu \\
&&b(a)=a \sigma(a)(\gamma +{1 \over 2} \partial_a {\sigma(a) \over
a}) \eea

\noindent with $\mu$, $\nu$ , $\gamma$ constant. The gauge-free
condition has therefore imposed the functional form of the drift
term. When the volatility function is fixed respectively to a
constant (Heston model), a  linear function  (geometric Brownian
model) and a quadratic function ($3/2$-model) in the volatility,
one obtains  the correct (mean-reversion) drift \footnote{In order
to obtain the correct number of parameters, one needs to apply a
change of local time $dt=\delta dt'$,
$dW_{1,2}=\sqrt{\delta}dW'_{1,2}$} (see Table 7)

\begin{table}[h]
\begin{center}
\begin{tabular}{|c|c|c|}
\hline name & $\sigma(a)$ & $SDE$ \\ \hline Heston &
$\sigma(a)=\eta$ & $dv=\sqrt{\delta}(2 v
\gamma+\eta(\eta-1))dt+2\eta \sqrt{\delta}\sqrt{v} dW_2$ \\ \hline
Geometric Brownian & $\sigma(a)=\eta a$ & $dv=\sqrt{\delta}(2 \eta
\gamma v^{3 \over 2}+\eta^2 v)dt+2 \sqrt{\delta}\eta v dW_2$
\\ \hline
$3/2$-model & $\sigma(a)=\eta a^2$ &
$dv=2\sqrt{\delta}\eta(\eta+\gamma)v^2 dt + 2\sqrt{\delta}\eta
v^{3 \over 2} dW_2$ \\ \hline
\end{tabular}
\caption{Examples of Gauge free stochastic volatility models with
$df={\delta}(\mu+\nu f)\sqrt{v} dW_1'$.}
\end{center}
\end{table}

\noindent The gauge transformation eliminating the connection is
then

\bea \Lambda(f,a)=-{\rho^{ff}  \over 2}ln({C(f)})+\alpha \rho^{fa}
\int^f_{} {df' \over C(f')}+(\rho^{ff} \alpha -{\rho^{af} \over 2}
\partial_f C) \int_{}^a {a' da' \over \sigma(a')} \eea

\noindent Finally, plugging the expression for $C(f)$ and $b(a)$
into (\ref{qqq}), we find that the function $Q$ is \footnote{$A$
and $B$ are two constants given by \beaa &&A={1 \over 2}(-{1 \over
2}\rho^{ff}
\partial_f C+\alpha \rho^{af})^2+{1 \over 2}\rho^{ff} C^2
\partial_f^2 ln(C) +\rho(-{1 \over 2}\rho^{ff} \partial_f C+\alpha
\rho^{af})(\alpha \rho^{ff}-{\rho^{af} \over 2}\partial_f C)+{1
\over 2}(\rho^{ff} \alpha -{\rho^{af} \over 2}\partial_f C)^2
\\
&&B=-{1 \over 2}(\rho^{ff} \alpha -\rho^{af} {\partial_f C \over
2}) \eeaa}

\bea Q=A a^2+B \sigma^2 \partial_a { a \over \sigma} \eea

\noindent The Black-Scholes equation for a Vanilla option
${\mathcal C}(a,f,\tau=T-t)$ (with strike $K$ and maturity $T$)
satisfied by the gauge transformed function ${\mathcal
C}'(a,f,\tau)=e^{\Lambda(f,a)}{\mathcal C}(a,f,\tau)$ is

\bea \partial_\tau {\mathcal C}'(a,f,\tau)=\Delta {\mathcal
C}'(a,f,\tau)+Q(a) {\mathcal C}'(a,f,\tau) \label{bs} \eea

\noindent with  the initial condition $ {\mathcal
C}'(a,f,\tau=0)=e^{\Lambda(f,a)}(f-K)^+$. In the coordinates
$q(f)=\int^f {df' \over C(f')}$ and $a$, the Laplace-Beltrami
operator is given by

\bea \Delta={a \sigma \over 2}( {a \over \sigma}\partial_q^2+2
\rho
\partial_{aq}+\partial_a{\sigma \over a}\partial_a) \eea

\noindent Applying a Fourier transformation according to $q$,
${\mathcal C}'(q,a,\tau)={\mathcal F}{\mathcal C}'(f,a,\tau)$, we
obtain

\bea \partial_\tau {\mathcal C}'(a,q,\tau)={a \sigma \over 2}(
-k^2{a \over \sigma}+2 i k \rho
\partial_{a}+\partial_a{\sigma \over a}\partial_a) {\mathcal
C}'(a,q,\tau)+Q(a){\mathcal C}'(a,q,\tau) \label{bs1} \eea

\noindent with the initial condition ${\mathcal
C}'(a,q,\tau=0)={\mathcal F}[e^{\Lambda(f,a)}(f-K)^+]$. Using a
spectral decomposition ${\mathcal C}'(a,q,\tau)=\sum_n
\phi_{nk}(a) ( {\mathcal C}'(a,q,\tau=0),\phi_{nk})e^{-E_{nk}
\tau}$ (with $(.,.)$ the scalar product on $L^2$), the
eigenvectors $\phi_{nk}(a)$ satisfy the equation

\bea -E_{nk}\phi_{nk}(a)={a \sigma \over 2}( -k^2{a \over
\sigma}+2 i k \rho
\partial_{a}+\partial_a{\sigma \over a}\partial_a)\phi_{nk}(a)+Q(a)\phi_{nk}(a)\label{bs2} \eea

\noindent This equation (\ref{bs2}) can be further simplified by
applying a {\it Liouville transformation} consisting in a gauge
transformation and a change of variable \cite{mil}

 \bea
\psi_{nk}(s)&=&({\sigma \over a})^{1 \over 2} e^{ik \rho \int^a{a'
\over \sigma(a')}da'} \phi_{nk}(a) \\
{ds \over da}&=&{\sqrt{2} \over \sigma(a)}\eea

\noindent $\psi_{nk}(s)$ satisfies a Schrodinger equation

\bea \psi_{nk}''(s)+(E_{nk}-J(s))\psi_{nk}(s)=0 \eea

\noindent with the scalar potential

\beaa J(s)=Q(a)-{k^2 a^2 \over 2}+ \frac{4\,a^4\,k^2\,{\rho }^2 -
3\,{\sigma (a)}^2 + a^2\,{\sigma '(a)}^2 +
    2\,a\,\sigma (a)\,\left( \sigma '(a) - a\,\sigma ''(a) \right) }{8\,a^2}
+{1 \over 2}\{a,s\}
    \eeaa

\noindent and where the curly bracket denotes the Schwarzian
derivative of $a$ with respect to $s$

\bea \{a,s\}=({ a''(s) \over a'(s)})'-{1 \over 2}({a''(s) \over
a'(s)})^2 \eea

\noindent The $2d$ partial differential equation corresponding to
our original KBS equation for our stochastic volatility model has
therefore been reduced via a change of coordinates and gauge
transformations to a Schrodinger equation with a scalar potential
$J(s)$. The stochastic volatility model is therefore solvable in
terms of hypergeometric functions if the potential $J(s)$ belongs
to the Natanzon class. The solution is given (in terms of the
eigenvectors $\psi_{nk}$) by

\bea {\mathcal C}(a,f,\tau)=e^{-\Lambda(a,f)}{\mathcal
F}^{-1}[\sum_n \psi_{nk}(s(a)) ({\mathcal
F}[e^{\Lambda(f,a)}(f-K)^+],\psi_{nk}(s(a)))e^{-E_{nk} \tau}] \eea

\noindent Let us examine classical examples of solvable stochastic
volatility model and show that the corresponding potentials $J(s)$
correspond to the Natanzon class (see Table 8).

\begin{table}[h]
\begin{center}
\begin{tabular}{|c|c|c|}
\hline name & potential & $J(s)$ \\ \hline

Heston & 3D osc. & $J(s)=\frac{-3 + 4\,B\,s^2\,\eta  + s^4\,{\eta
}^2\,\left( 2\,A + k^2\,\left( -1 + {\rho }^2 \right) \right)
}{4\,s^2}$ \\ \hline Geometric Brownian & Morse &
$J(s)=\frac{-{\eta }^2 + 4\,e^{{\sqrt{2}}\,s\,\eta }\,\left( 2\,A
+ k^2\,\left( -1 + {\rho }^2 \right)  \right) }{8}$ \\ \hline
$3/2$-model & Coulombian & $J(s)=\frac{8\,A + \eta \,\left( -8\,B
+ \eta  \right)  + 4\,k^2\,\left( -1 + {\rho }^2 \right)
}{4\,s^2\,{\eta }^2}$ \\ \hline
\end{tabular}
\caption{Stochastic volatility models and potential $J(s)$}
\end{center}
\end{table}

\noindent We present here a new example of solvable stochastic
volatility model which corresponds to the Posh-Teller I potential.

\begin{exa}[Posh-Teller I]
For a volatility function given by $ \sigma(a)=\alpha  +\eta a^2
$, the potential is given by a Posh-Teller I potential \footnote{
$csc(z) \equiv {1 \over sin(z)}$ and $sec(z) \equiv {1 \over
cos(z)}$} \beaa J(s)=\frac{\alpha \left( 4\left( -2A + k^2 +
4B\eta - k^2{\rho }^2 \right)  -
      3{\eta }^2{\csc (\frac{s{\sqrt{\alpha }}{\sqrt{\eta }}}{{\sqrt{2}}})}^2 +
      \left( 8A + \eta \left( -8B + \eta  \right)  + 4k^2\left( -1 + {\rho }^2 \right)  \right)
       {\sec (\frac{s{\sqrt{\alpha }}{\sqrt{\eta }}}{{\sqrt{2}}})}^2 \right) }{8\eta
       }\eeaa
\end{exa}

\section{Conclusion}
We have shown how to use supersymmetric methods to generate new
solutions to the Kolmogorov \& Black-Scholes equation (KBS) for
one-dimensional diffusion processes. In particular, by applying a
supersymmetric transformation on the Natanzon potential, we have
generated a hierarchy of new solvable processes. Then, we have
classified the stochastic volatility models which admit a flat
Abelian connection (with one Killing vector). The two-dimensional
KBS equation has been converted into a Schrodinger equation with a
scalar potential. The models for which the scalar potential
belongs to the Natanzon class are solvable in terms of
hypergeometric functions. This is the case for the Heston model,
the geometric brownian model and the $3/2$-model. A new solution
with a volatility of the volatility $ \sigma(a)=\alpha  +\eta a^2
$, corresponding to the Posh-Teller I, has been presented.

\section*{Acknowledgements}
I would like to thank Dr. S. Ribault for useful comments and
discussions and Dr. C. Waite for a careful reading.

\end{document}